\documentclass[reprint,superscriptaddress,preprintnumbers,nofootinbib,amsmath,amssymb,aps]{revtex4-2}

\usepackage{graphicx}% Include figure files
\usepackage{dcolumn}% Align table columns on decimal point
\usepackage{bm}% bold math
\usepackage{natbib}
\usepackage{booktabs}
\usepackage{enumitem}
\usepackage{amssymb}    %Allows use math symbols
\usepackage{color}         %Allows {\color{red} Hello World} or \colorbox{blue}{Hello World}
\usepackage{graphicx}     %Allows\includegraphics{figure.pdf}
\usepackage{mathrsfs} %Allows \mathscr{HELLO}
\usepackage{hyperref} %Automatically links \label and \ref commands; Always load last
\hypersetup{colorlinks=true,linkcolor=blue,citecolor=blue}
\usepackage{ upgreek } % for uptau
\usepackage{color,xcolor,fancybox,epsf,rotating,colordvi}

\newcommand{\SLASH}[2]{\makebox[#2ex][l]{$#1$}/}
\newcommand{\eslash}{\SLASH{E}{.2}}

\newcommand{\GeV}{~\rm GeV}
\newcommand{\TeV}{~\rm TeV}

\newcommand{\abm}{{~\rm ab}^{-1}}

\begin{document}

% \preprint{APS/123-QED}

% \title{Probing Heavy Higgs Bosons with a 100 TeV Hadron Collider \\ in the Semi-Constrained NMSSM}% Force line breaks with \\
\title{Exploring Heavy Higgs Bosons at a 100 TeV Hadron Collider \\ within the Semi-Constrained NMSSM}
\author{Kun Wang}
\email[]{kwang@usst.edu.cn}
\affiliation{College of Science, University of Shanghai for Science and Technology,  Shanghai 200093,  China}

\author{Pengfu Tian}
\affiliation{School of Physics and Technology, Wuhan University, Wuhan 430072, China}

\author{Jingya Zhu}
\email[]{zhujy@henu.edu.cn} 
%\email[Corresponding author:]{zhujy@henu.edu.cn} 
\affiliation{School of Physics and Electronics, Henan University,  Kaifeng 475004, China}

% \collaboration{CLEO Collaboration}%\noaffiliation

\date{\today}% It is always \today, today,

\begin{abstract}
In this study, we explore the detectability of heavy Higgs bosons in the $pp \to b\bar{b}H/A \to b\bar{b}t\bar{t}$ channel at a 100 TeV hadron collider within the semi-constrained Next-to-Minimal Supersymmetric Standard Model (NMSSM). 
We calculate their production cross sections and decay branching ratios, comparing these with simulation results from existing reference. 
We focus on the heavy, doublet-dominated CP-even Higgs $H$ and CP-odd Higgs $A$, with mass limits set below 10 TeV to ensure detectability. 
We find that at a collider with 3 ab$^{-1}$ of integrated luminosity, the potential for detecting heavy Higgs bosons varies significantly with their mass and $\tan\beta$. 
Heavy Higgs bosons below 2 TeV are within the testable range, while those heavier than 7 TeV fall below the exclusion and discovery thresholds, rendering them undetectable. 
For masses between 2 and 7 TeV, heavy Higgs bosons with $\tan\beta$ less than 20 can be detected, whereas those with $\tan\beta$ greater than 20 are beyond the current discovery or exclusion capabilities.
\end{abstract}

\maketitle
\newpage

% \tableofcontents
% \newpage

\section{\label{sec:Introduction}Introduction}

It has been widely acknowledged for an extended period that the Standard Model (SM) of particle physics, despite its remarkable success in explaining a vast array of phenomena, fails to provide a complete description of the fundamental aspects of the universe. 
Consequently, the search for new physics beyond the SM (BSM) has become a crucial direction in modern physics research.

The discovery of a Higgs boson with properties consistent with SM predictions at the Large Hadron Collider (LHC) in 2012  significantly bolstered our understanding of the SM \cite{ATLAS:2012yve, CMS:2012qbp,ATLAS:2022vkf,CMS:2022dwd}. 
Nevertheless, this milestone also intensified the debate regarding the possible existence of additional Higgs bosons. 
These scalar particles are theoretically predicted by a large number of natural BSM models, which aim to address the limitations and unresolved questions of the SM. 
Among these are the Minimal Supersymmetric Standard Model (MSSM) \cite{Medina:2017bke, Bechtle:2016kui, Barman:2016kgt, Carena:2015uoe, Djouadi:2013vqa, Bechtle:2012jw,Baer:2022smj,Chattopadhyay:2022ecq,Baer:2021qxa,Bahl:2020kwe,Adhikary:2018ise,BhupalDev:2014bir}, 
the Next-to-MSSM (NMSSM) \cite{Ma:2020mjz, Shang:2020uog, Cao:2016uwt, Wang:2015omi, Carena:2015moc, Bomark:2014gya, Cao:2014kya, King:2014xwa, Baglio:2013iia, Cao:2012fz, Cao:2012yn, Cao:2013gba,Heng:2023wqf,Arganda:2021qgi,Li:2023kbf,Wang:2021lwi,Li:2023tlk}, 
the Minimal Dilaton Model (MDM) \cite{Cao:2016udb, Agarwal:2016gxe, Cao:2013cfa, Liu:2018ryo,Wang:2024bkg},   and the Little Higgs Models (LHM) \cite{Han:2013ic, Cheung:2018ljx}, among others.
The experimental search for heavy neutral Higgs bosons has focused on decay channels such as $\tau^+\tau^-$ \cite{CMS:2018rmh, ATLAS:2017eiz}, $t\bar{t}$, $b\bar{b}$, $\mu^+\mu^-$, $ZZ$, $WW$, $hh$, and $hV$, which are reviewed in \cite{ParticleDataGroup:2020ssz, ATLAS:2018sbw}.
And recently evidence for a new scalar around 151 GeV was accumulated with significances of $4.3\sigma$ local and $3.9\sigma$ global \cite{Crivellin:2021ubm}, subsequently revised to $4.1\sigma$ locally and $3.5\sigma$ globally \cite{Fowlie:2021ldv}.

The exploration of new physics phenomena, especially the detection of heavy Higgs bosons, requires colliders with higher energy. 
Future high-energy colliders, designed to exceed the capabilities of current facilities, will be able to examine these heavy particles.
Notably, the Future hadron-hadron Circular Collider (FCC-hh) at CERN \cite{Gomez-Ceballos:2013zzn} and the Super-$pp$-Collider (SppC) \cite{Gao:2021bam, CEPCStudyGroup:2018ghi} in China are among the most ambitious projects in this direction. 
Both initiatives aim to construct a 50-100 TeV $pp$ collider \cite{Hinchliffe:2015qma}, promising a significant leap in the energy frontier and potentially uncovering phenomena beyond the SM.
Moreover, the concept of a multi-TeV muon collider presents an innovative approach to high-energy physics experiments \cite{Bartosik:2020xwr, NeutrinoFactory:2002azy}. 
Extensive research has been conducted on the detection of heavy Higgs bosons at future colliders. 
In particular, the studies described in Ref. \cite{Hajer:2015gka} examined the $pp\to b\bar{b}H/A\to b\bar{b}\tau\tau$ and $pp\to b\bar{b}H/A\to b\bar{b}t\bar{t}$ channels at a 100 TeV $pp$ collider, and proposed pushing the exclusion limits for heavy Higgs searches up to $M_H \sim 10\TeV$, with exceptions in regions of low $\tan\beta$.
Furthermore, the analysis in Ref. \cite{Baer:2021qxa} explored the $pp\to H/A \to \tilde{\chi}_1^0\tilde{\chi}_2^\mp$ process, revealing the $4\ell+\eslash$ signal at a 100 TeV hadron collider, demonstrating its ability to probe new supersymmetric model sectors. 
In addition, Ref. \cite{Han:2021udl} explored the potential of a multi-TeV muon collider to discover heavy Higgs bosons within Two Higgs Doublet Models (2HDMs)  and assess the discriminative power among different 2HDM types.

In the current study, we extend the investigation initiated in our previous work on heavy Higgs bosons within the framework of the semi-constrained NMSSM (scNMSSM) \cite{Wang:2020tap}. 
The NMSSM incorporates an additional singlet superfield to the MSSM, thereby enriching the Higgs and neutralino sectors. 
Our analysis focuses on the computational evaluation of production cross sections and decay branching ratios for these heavy Higgs bosons. 
Through these calculations, we aim to provide a comprehensive understanding of the behavior and detectability of heavy Higgs bosons within the scNMSSM. 
Furthermore, we delve into the exploration of the discovery potential of these heavy Higgs bosons through the $pp\to b\bar{b}H/A\to b\bar{b}t\bar{t}$ channel at a future 100 TeV collider. 
The selection of a 100 TeV collider is driven by its exceptional ability to achieve the high energy levels required for producing such massive particles, thus opening up new avenues for their discovery.

The remainder of this manuscript is organized as follows. 
In Sec.~\ref{sec:model}, we provide a brief overview of the scNMSSM, outlining its fundamental aspects and theoretical significance. 
In Sec.~\ref{sec:res}, we present a detailed account of our computational methodology, followed by a comprehensive discussion of the results obtained from our analysis. 
In Sec.~\ref{sec:con}, we conclude the paper by summarizing the main results and their implications for future research in this area.

\section{The semi-constrained NMSSM}
\label{sec:model}

The NMSSM extends the MSSM by introducing an additional singlet superfield, denoted as $\hat{S}$, where the superpotential of the $\mathbb{Z}_3$-symmetric NMSSM is defined as:
\begin{equation}
    W_{\mathrm{NMSSM}} = W_{\mathrm{MSSM}}|_{\rm \mu=0} + \lambda\hat{S} \hat{H}_{u} \cdot \hat{H}_{d} + \frac{\kappa}{3} \hat{S}^{3} \,, 
\end{equation}
where $W_{\mathrm{MSSM}}|_{\rm \mu=0}$ is the superpotential of the MSSM without the $\mu$-term, $\lambda$ and $\kappa$ are coupling constants, $\hat{H}_{u}$ and $\hat{H}_{d}$ are the doublet Higgs superfields, and $\hat{S}$ is the added singlet superfield.
After electroweak symmetry breaking, the singlet scalar's vacuum expectation value (VEV), denoted $v_{s}$, dynamically generates the massive $\mu$-term \cite{Ellwanger:2009dp, Maniatis:2009re}
\begin{equation}
    \mu \equiv \lambda v_s \, .
\end{equation}
Concurrently, the scalar components $H_{u}$ and $H_{d}$ also attain VEVs, labeled $v_{u}$ and $v_{d}$, respectively.
This leads to the introduction of a new parameter $\tan \beta$, defined as
\begin{align}
    \tan\beta \equiv v_u/v_d  \, ,
\end{align}
where the sum of their squares is $v_u^2 + v_d^2 = v^2 =  (174 \GeV)^2 $.

The NMSSM introduces specific soft SUSY breaking terms, distinct from those in the MSSM, as given by:
\begin{align}
    - {\cal L}_{\mathrm{N M S S M}}^{\mathrm{s o f t}}=&-{\cal L}_{\mathrm{M S S M}}^{\mathrm{s o f t}} |_{\mu=0}+m_{S}^{2} | S |^{2} \nonumber \\ 
    &+\lambda A_{\lambda} S H_{u} \cdot H_{d}  +\frac{1} {3} \kappa A_{\kappa} S^{3}+\mathrm{h. c.} \,, 
\end{align}
where ${\cal L}_{\mathrm{MSSM}}^{\mathrm{soft}} |_{\mu=0}$ denotes the MSSM's soft SUSY breaking terms with the $\mu$ parameter set to zero. The symbols $H_{u}$ and $H_{d}$ refer to the scalar components of the Higgs doublets, $A_{\lambda}$ and $A_{\kappa}$ represent the trilinear coupling constants with mass dimension, and $m_{S}$ is the mass of the singlet scalar field.

In the scNMSSM, the Higgs sector is allowed to deviate from universality at the Grand Unified Theory (GUT) scale, a characteristic also known as the NMSSM with non-universal Higgs masses. 
Specifically, the soft masses for the Higgs fields, $m_{H_{u}}^2$, $m_{H_{d}}^2$, and $m_{S}^2$, can differ from $M_0^2 + \mu^2$. 
Furthermore, the trilinear coupling constants $A_{\lambda}$ and $A_{\kappa}$ may vary independently from $A_0$. 
Consequently, the parameter space of the scNMSSM is defined by nine parameters:
\begin{align}
    \lambda, \ \kappa, \ \tan \beta, \ \mu, \ A_{\lambda}, \ A_{\kappa}, \ A_{0}, \ M_{1/2}, \ M_{0} \, .
\end{align}
Here, $M_{1/2}$ and $M_{0}$ represent the universal sfermion mass and the universal gaugino mass, respectively, while $A_{0}$ denotes the universal trilinear coupling constant in the sfermion sector.

The Higgs sector within the NMSSM is predicted to contain three CP-even Higgs bosons, two CP-odd Higgs bosons, and a pair of charged Higgs bosons. 
For convenience, the scalar components of the superfields $H_u$, $H_d$, and $S$ are often rotated so that they can be represented as
\begin{align}
    H_{1} &= \cos \beta H_{u} + \varepsilon \sin \beta H_{d}^{*} = \left( \begin{matrix} H^{+} \\ \frac{S_{1} + i P_{1}^{\,}}{\sqrt{2}} \end{matrix} \right), \\ 
    H_{2} &= \sin \beta H_{u} - \varepsilon \cos \beta H_{d}^{*} = \left( \begin{matrix} G^{+} \\ v + \frac{S_{2} + i G^{0}}{\sqrt{2}} \end{matrix} \right), \\
    H_{3} &= S = v_{s} + \frac{S_{3} + i P_{2}}{\sqrt{2}},
\end{align}
where $\varepsilon = \left( \begin{smallmatrix} 0 & 1 \\ -1 & 0 \end{smallmatrix} \right)$, 
and $S_1$, $S_2$, and $S_3$ create the CP-even basis, while $P_1$ and $P_2$ establish the CP-odd one.
$H_2$ is identified as the SM-like Higgs, $H_1$ represents a new Higgs doublet field, and $H_3$ introduces a new singlet field.

% The CP-even Higgs mass eigenstate is mix by $(S_1, S_2, S_3)$, so the mass matrix $M_S^2$ is given by 
The CP-even Higgs mass matrix $M_S^2$ in the basis  $(S_1, S_2, S_3)$ is given by \cite{Carena:2015moc,Miller:2003ay}:
\begin{align}
    M_{S, 11}^{2}&= M_{A}^{2}+( m_{Z}^{2}-\lambda^{2} v^{2} ) \mathrm{s i n}^{2} 2 \beta, \label{eq:m11}\\ 
    M_{S, 12}^{2}&=  {-} \frac{1} {2} ( m_{Z}^{2} \!-\! \lambda^{2} v^{2} ) \mathrm{s i n} 4 \beta, \\ 
    M_{S, 13}^{2}&=  {-} \left( \frac{M_{A}^{2}} {2 \mu/ \mathrm{s i n} 2 \beta} \!+\! \kappa v_{s} \right) \lambda v \mathrm{c o s} 2 \beta, \\
    M_{S, 22}^{2}&=m_{Z}^{2} \operatorname{c o s}^{2} 2 \beta+\lambda^{2} v^{2} \operatorname{s i n}^{2} 2 \beta, \\
    M_{S, 23}^{2}&=2 \lambda\mu v \left[ 1-\left( \frac{M_{A}} {2 \mu/ \operatorname{s i n} 2 \beta} \right)^{2}-\frac{\kappa} {2 \lambda} \operatorname{s i n} 2 \beta\right], \\
    M_{S, 33}^{2}&=\frac{1} {4} \lambda^{2} v^{2} \left( \frac{M_{A}} {\mu/ \operatorname{s i n} 2 \beta} \right)^{2}+\kappa v_{s} A_{\kappa}+4 ( \kappa v_{s} )^{2} \nonumber \\
    & \quad -\frac{1} {2} \lambda\kappa v^{2} \operatorname{s i n} 2 \beta,
\end{align}
where $M_A$ is defined as:
\begin{align}
    M_A = \frac{2 \mu( A_{\lambda} {+} \kappa v_{s} )} {\operatorname{s i n} 2 \beta}.
\end{align}

% The CP-odd Higgs mass eigenstate is mix by $(P_1, P_2)$, so the mass matrix $M_P^2$ is given by 
The CP-odd Higgs mass matrix $M_P^2$ in the basis  $(P_1, P_2)$ is given by: 
\begin{align}
    M_{P, 1 1}^{2}&=M_{A}^{2} \, , \label{eq:mp11} \\  
    M_{P, 1 2}^{2}&=\lambda v ( A_{\lambda}-2 \kappa v_{s} ) \, ,\\ 
    M_{P, 2 2}^{2}&=\lambda( A_{\lambda}+4 \kappa v_{s} ) \frac{v_{u} v_{d}} {v_{s}}-3 \kappa v_{s} A_{\kappa} \,.  
\end{align}

% Three CP-even mass eigenstates $h_1$, $h_2$, and $h_3$ ($m_{h_1}<m_{h_2}<m_{h_3}$) are mixed from $(S_1, S_2, S_3)$, and two CP-odd mass eigenstates $a_1$ and $a_2$ ($m_{a_1}<m_{a_2}$) are mixed from $(P_1, P_2)$, it can be writeen as
% \begin{align}
%     \begin{pmatrix} h_{1} \\h_{2} \\h_{3} \\ \end{pmatrix} &= S_{ij}\begin{pmatrix}S_{1} \\S_{2} \\S_{3} \\ \end{pmatrix},  \\
%     \begin{pmatrix}a_{1} \\a_{2} \\ \end{pmatrix} &= P_{ij}\begin{pmatrix}P_{1} \\P_{2} \\ \end{pmatrix},
% \end{align}
% where $S_{ij}$ and $P_{ij}$ are the mix matrix, it can diagnal the mass matrix $M_{S}^{2}$ and $M_{P}^{2}$. 

Three CP-even mass eigenstates $h_1$, $h_2$, and $h_3$ ($m_{h_1}<m_{h_2}<m_{h_3}$) are derived from the mixture of $(S_1, S_2, S_3)$, and two CP-odd mass eigenstates $a_1$ and $a_2$ ($m_{a_1}<m_{a_2}$) are derived from $(P_1, P_2)$. This can be represented as:
\begin{align}
    \begin{pmatrix} h_{1} \\h_{2} \\h_{3} \\ \end{pmatrix} &= S_{ij}\begin{pmatrix}S_{1} \\S_{2} \\S_{3} \\ \end{pmatrix},  \\
    \begin{pmatrix}a_{1} \\a_{2} \\ \end{pmatrix} &= P_{ij}\begin{pmatrix}P_{1} \\P_{2} \\ \end{pmatrix},
\end{align}
where $S_{ij}$ and $P_{ij}$ are the mixing matrices that diagonalize the mass matrices $M_S^2$ and $M_P^2$, respectively.

Among the three CP-even Higgs bosons ($h_i$, where $i=1,2,3$), the $125 \GeV$ SM-like Higgs could be either $h_1$ or $h_2$, both of which are predominantly doublet-dominated scalars. 
The remaining CP-even Higgs bosons include another doublet-dominated and a singlet-dominated scalar. 
For the two CP-odd Higgs bosons ($a_i$, where $i=1,2$), one is doublet-dominated, and the other is singlet-dominated.
The singlet-dominated Higgs boson rarely couples to fermions because the singlet $ S $ interacts only with the Higgs sector. 
This property makes it difficult to detect at the LHC.
In contrast, the doublet-dominated Higgs boson couples to fermions, which facilitates its detection. 
Our study focuses only on the heavy, doublet-dominated CP-even Higgs $H$ and CP-odd Higgs $A$, because of their detectability. 
The couplings to up/down-type fermions of these heavy Higgs bosons, $H$ and $A$, are defined as follows:
\begin{align}
    C_{Huu}^{SUSY} &= i \frac{m_{u}} {v}  \cot \beta \\
    C_{Hdd}^{SUSY} &= i \frac{m_{d}} {v}  \tan \beta  \\
    C_{Auu}^{SUSY} &= \frac{m_{u}} {v} \cot \beta \gamma_{5} \\
    C_{Add}^{SUSY} &= \frac{m_{d}} {v} \tan \beta \gamma_{5} 
\end{align}
The reduced couplings of these heavy Higgs bosons, $H$ and $A$, are defined as follows:
\begin{align}
    C_{Huu}=C_{Huu}^{SUSY}/C_{Huu}^{SM} &=  \cot \beta \\
    C_{Auu}=C_{Auu}^{SUSY}/C_{Auu}^{SM} &= \cot \beta \\
    C_{Hdd}=C_{Hdd}^{SUSY}/C_{Hdd}^{SM} &= \tan \beta \\
    C_{Add} = C_{Add}^{SUSY}/C_{Add}^{SM} &= \tan \beta
\end{align}

Furthermore, it is observed that when $ \tan \beta $ is significantly larger than 1 ($ \tan \beta \gg 1 $), $ M_{S, 11}^2 $ closely approximates $ M_A^2 $. 
In addition, the Higgs bosons $ H $ and $ A $ become degenerate, meaning they have the same mass and exhibit identical couplings to quarks.

\section{Results and Discussions}
\label{sec:res}

% In this work, we focus on the heavy, doublet-dominated CP-even Higgs $H$ and CP-odd Higgs $A$ within the scNMSSM, and investigate their detectability at the 100 TeV Hadron Collider. 
In this study, we explore the detectability of the heavy, doublet-dominated CP-even Higgs ($H$) and CP-odd Higgs ($A$) in the scNMSSM, at a 100 TeV Hadron Collider. 
We set the upper mass limit for the Higgs at 10 TeV, represented as:
\begin{align}
    h_i, a_j < 10 \TeV \quad \text{for } i = 1, 2, 3;~j=1, 2.
\end{align}
Therefore, we consider the relevant parameter space in the scNMSSM as follows:
\begin{gather*}
0.0<\lambda<0.7, \quad  |\kappa|<0.7, \quad 1<\tan\beta <60 \, , \\
0.0<\mu, \,\, M_0, \,\, M_{1/2} < 10\TeV \, , \\
|A_0|,\,\, |A_\lambda|,\,\, |A_\kappa| < 10 \TeV \, .
\end{gather*}

We use the package \textsf{NMSSMTools-6.0.2} \cite{Ellwanger:2004xm, Ellwanger:2005dv, Ellwanger:2006rn, Das:2011dg} to scan the parameter space and calculate relevant quantities, considering the following constraints:
(i) Theoretical constraints including vacuum stability and without Landau pole below the GUT scale\cite{Ellwanger:2004xm, Ellwanger:2005dv}.
(ii) Flavor constraints from rare B-meson decays and $D$-meson mass differences\cite{Tanabashi:2018oca,  Aaij:2012nna, Lees:2012xj, Lees:2012ym}.
(iii) A 123–127 GeV Higgs boson with signal predictions that are globally consistent with LHC Higgs data \cite{Aad:2019mbh, Sirunyan:2018koj, Khachatryan:2016vau,ATLAS:2022vkf,CMS:2022dwd,CMS:2018uag,ATLAS:2016neq}.
(iv) Constraints from searches for additional Higgs bosons and exotic decays of the SM-like Higgs, using \textsf{HiggsBounds-5.5.0}, including a limit of $10.7\%$ on invisible Higgs decay \cite{Bechtle:2015pma, Bechtle:2013wla, Bechtle:2013gu, Bechtle:2011sb, Bechtle:2008jh,ATLAS:2023tkt}.
(v) Upper bounds on the dark matter relic density from WMAP/Planck \cite{Hinshaw:2012aka, Ade:2013zuv}.
(vi) Direct dark matter search constraints from XENON1T \cite{Aprile:2018dbl, Aprile:2019dbj}, PICO-60 \cite{Amole:2019fdf}, PandaX-4T \cite{PandaX:2022xas,PandaX-4T:2021bab} and LUX-ZEPLIN \cite{LZ:2022lsv}.
(vii) Constraints from direct SUSY searches at the LHC and LEP, using the package \textsf{SModelS-v2.2.1} \cite{Kraml:2013mwa, Ambrogi:2017neo, Ambrogi:2018ujg, Dutta:2018ioj, Buckley:2013jua, Sjostrand:2006za, Sjostrand:2014zea}.

% \begin{enumerate}[label=(\roman*)][label={(\arabic*)}]
%     \item 
%     \item 
% \end{enumerate}

\begin{figure*}[!htbp]
\centering
\includegraphics[width=1\textwidth]{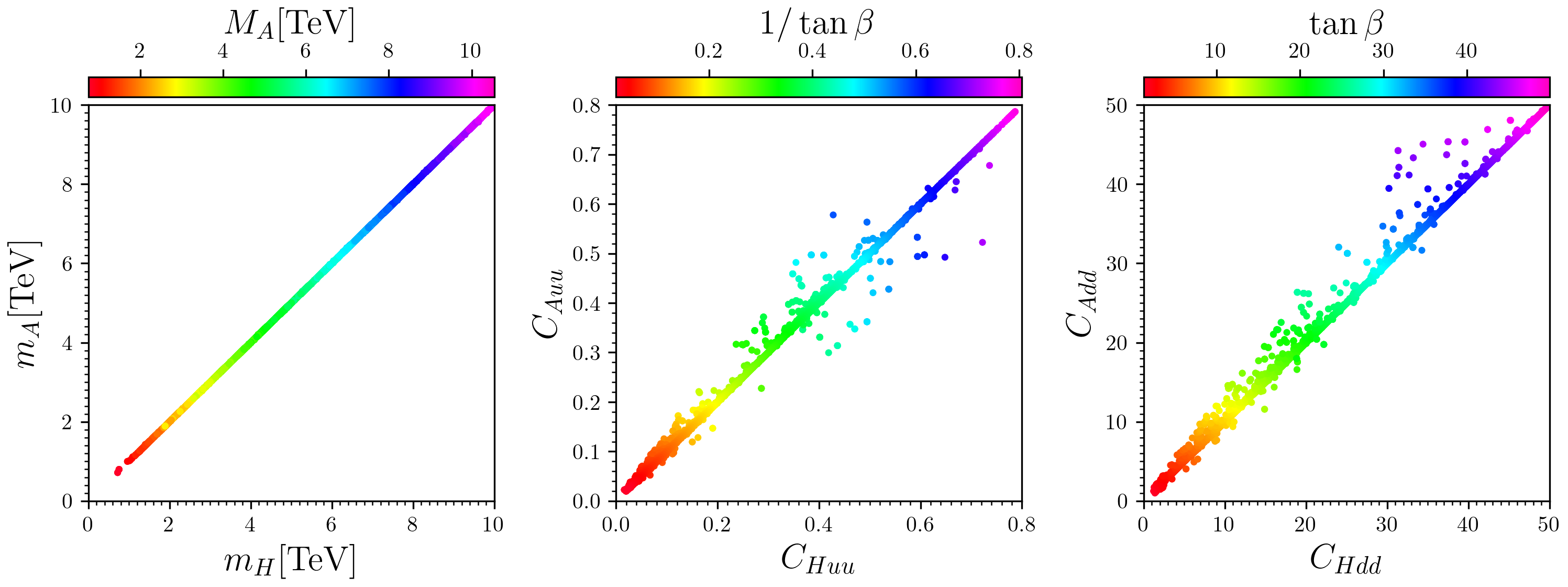}
\vspace{-0.6cm}
\caption{
Surviving samples are shown in the planes of $m_A$ versus $m_H$ (left), reduced coupling  $C_{Auu}$  versus $C_{Huu}$ (middle), and reduced coupling $C_{Add}$ versus $C_{Hdd}$ (right).
From left to the right, the colors represent $M_A$, $1/\tan \beta$, and $\tan \beta$ respectively.
Samples with larger values of $\tan \beta$ are plotted on top of those with smaller values.
}
\label{fig:1}
\end{figure*}

For the samples satisfying the above theoretical and experimental constraints, we observe the following properties:
\begin{itemize}
    \item The squarks of the first two generations are heavier than $2.2\TeV$, with the lightest squark, $\tilde{t}_1$, exceeding $1\TeV$.
    \item The third-generation sleptons can be as light as approximately $170\GeV$.
    \item The glugino mass exceeds $2\TeV$. Consequently, given the universal gaugino mass condition at the GUT scale, the bino and wino masses are more than $340\GeV$ and $620\GeV$, respectively.
    \item The mass range for the lightest neutralino varies from $4\GeV$ to $4\TeV$, typically dominated by bino and singlino compositions, with some higgsino admixture.   
    \item In the Higgs sector, we categorize the samples into two types:
\begin{itemize}
\item $h_1$ is the $125 \GeV$ SM-like Higgs.
$h_2$ and $h_3$ are heavy CP-even Higgs bosons, while $a_1$ and $a_2$ are heavy CP-odd Higgs bosons.
\item $h_2$ is the $125 \GeV$ SM-like Higgs. 
The light CP-even Higgs $h_1$ and the light CP-odd Higgs $a_1$  are typically singlet-dominant.
The heavy CP-even Higgs $h_3$ and the light CP-odd Higgs $a_2$ are typically doublet-dominant.
\end{itemize}
\end{itemize}

In this study, we focus on doublet-dominant heavy Higgs due to the difficulty of detecting singlet-dominant Higgs. 
There are two types to consider. In the first type, $h_1$ resembles the SM-like Higgs, with either $h_2$ or $h_3$ being doublet-dominant, and the same holds for $a_1$ and $a_2$, where one of them is doublet-dominant.
We label the heavy CP-even and CP-odd doublet-dominant Higgs as $H$ and $A$, respectively. 
In the second type, $h_2$ acts as the SM-like Higgs, with $h_3$ and $a_2$ typically being doublet-dominant, also denoted as $H$ and $A$. 
Thus, $H$ and $A$ represent the heavy CP-even and CP-odd doublet-dominant Higgs in subsequent discussions.
For the heavy Higgs $H$ and $A$, which have masses ranging from $0.6\TeV$ to $10\TeV$, we calculate their production cross sections and decay branching ratios. 
We also compare the $pp\to b\bar{b}H\to b\bar{b}t\bar{t}$ signal with simulation results found in Ref. \cite{Hajer:2015gka}.

In Fig. \ref{fig:1}, we show the mass and reduced coupling of the heavy doublet-dominated Higgs bosons $H$ and $A$ in the scNMSSM.
The following observations can be made from these figures:
\begin{itemize}
\item In the left panel, the surviving samples are plotted on the $m_A$ versus $m_H$ plane, with colors indicating $M_A$. 
It is observed that $H$ and $A$ have nearly identical masses, approximately equal to the parameter $M_A$. 
This similarity arises because, according to Eq.~\ref{eq:mp11}, $P_1$ is the CP-odd doublet-dominated Higgs, and since $A$ is also denoted as the CP-odd doublet-dominated Higgs, it follows that $m_A \approx M_A$. 
Furthermore, $S_1$ is the CP-even doublet-dominated Higgs, labeled here as $H$. From Eq.~\ref{eq:m11}, when $\tan\beta \gg 1$, it is derived that $m_H \approx M_A$.
And the mass of $H$ and $A$ is between $0.6\TeV$ and $10\TeV$.

\item In the middle panel, the surviving samples are displayed on the plane of reduced coupling with up-type fermions for $A$ versus $H$, with the colors indicating $1/\tan \beta$. 
It is evident that for most samples, $C_{Auu}$ and $C_{Huu}$ are approximately equal to $1/\tan \beta$. 
This approximation arises because the doublet components of $H$ and $A$ are neither exactly equal nor exactly equal to 1. 
Furthermore, the values of the reduced couplings $C_{Huu}$ and $C_{Auu}$ range from 0 to 0.8.

\item In the right panel, the surviving samples are plotted on the plane of reduced coupling with down-type fermions for $A$ versus $H$, with colors representing $\tan \beta$. 
The results are similar to those in the middle panel; for most samples, $C_{Add}$ and $C_{Hdd}$ approximate $\tan \beta$. 
This approximation is also due to the doublet components of $H$ and $A$ not being exactly equal or exactly equal to 1. 
Additionally, the values of the reduced couplings $C_{Hdd}$ and $C_{Add}$ vary from 0 to 50.

\end{itemize}

\begin{figure*}[!htbp]
\centering
\includegraphics[width=1\textwidth]{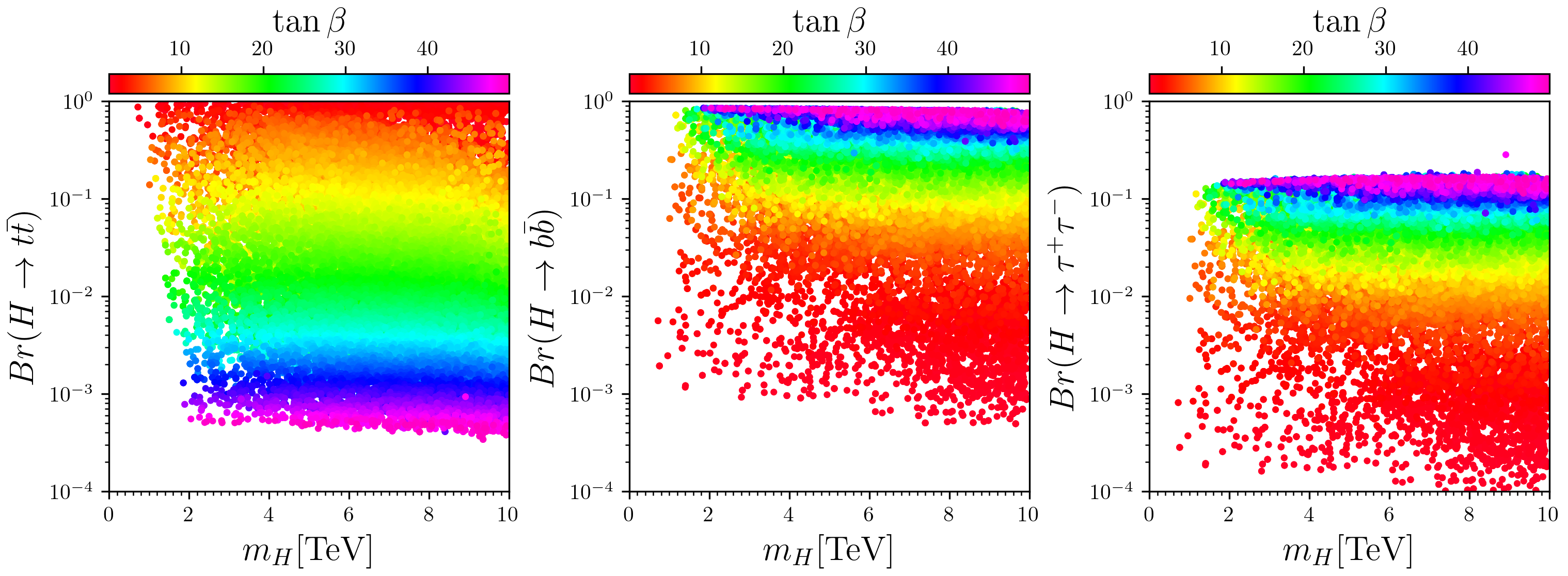}
\includegraphics[width=0.67\textwidth]{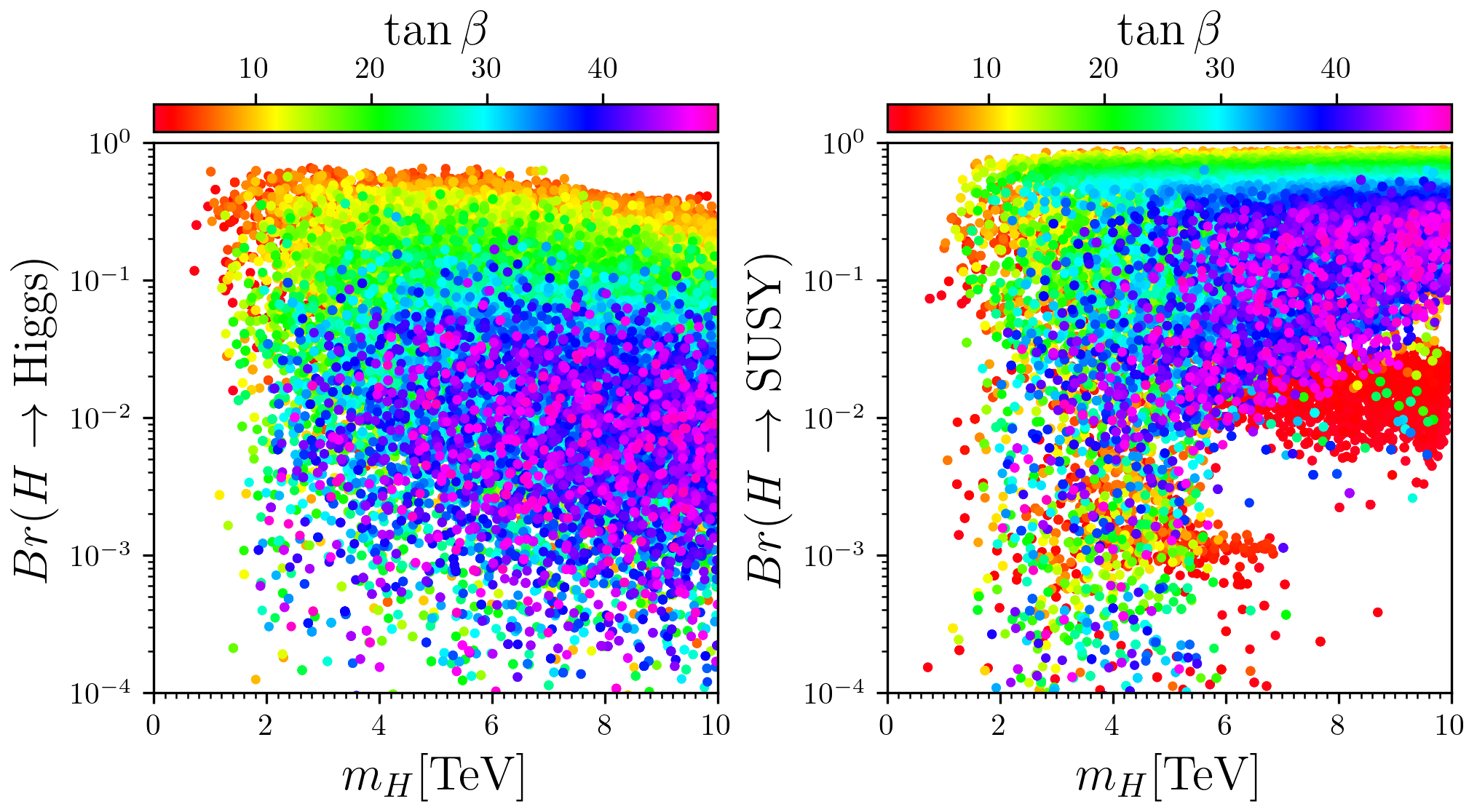}
% \vspace{-0.6cm}
\caption{
Surviving samples in the planes of branching ratios versus $m_H$, with colors representing $\tan\beta$. 
The branching ratios pertain to the decays of the heavy CP-even Higgs $H$ into $t\bar{t}$, $b\bar{b}$, $\tau^+ \tau^-$,  all possible lighter Higgs bosons, and all possible SUSY particles, respectively. 
Samples with larger values of $\tan \beta$ are plotted on top of those with smaller values.
}
\label{fig:2}
\end{figure*}

In Fig. \ref{fig:2} we show the decay properties of the CP-even doublet-dominated heavy Higgs $H$ in the scNMSSM, with colors representing $\tan \beta$. 
Since both heavy Higgs $A$ and $H$ are doublet-dominated, their couplings to fermions show very little difference. 
The only difference is that $Br(A\to VV/\tilde{V}\tilde{V})=0$. 
However, this difference is minimal due to the small coefficient $C_{HVV}$.
As a consequence, the decay properties of the CP-odd doublet-dominated heavy Higgs $A$ are very similar to those of $H$; therefore, the plot for $A$ is omitted. 
The following observations can be drawn from these plots:
\begin{itemize}
\item The dominant branching ratios consistently arise from the decays to $t\bar{t}$, $b\bar{b}$ and SUSY particles, with these branching ratios reaching values close to 1.

\item In the upper left panel, the branching ratio of $H$ to $t\bar{t}$ is inversely proportional to $\tan\beta$; that is, the smaller the $\tan\beta$, the larger the branching ratio $Br(H \to t\bar{t})$. 
This relationship is due to $C_{Huu}$ being directly proportional to $1/\tan\beta$. 
Consequently, when $\tan\beta < 10$, the branching ratio $Br(H \to t\bar{t})$ exceeds 0.2. 
As $\tan\beta$ approaches 1, the branching ratio $Br(H \to t\bar{t})$ tends towards 1.

\item In the upper middle and right panels, the branching ratios of $H$ to $b\bar{b}$ and $\tau^+ \tau^-$ are proportional to $\tan\beta$; specifically, the larger the $\tan\beta$, the higher the branching ratios $Br(H \to b\bar{b})$ and $Br(H \to \tau^+ \tau^-)$. 
This proportionality is due to $C_{Hdd}$ being directly proportional to $\tan\beta$. 
Additionally, when $\tan\beta$ exceeds 40, the maximum branching ratio $Br(H \to b\bar{b})$ can reach up to 0.8, while the maximum branching ratio $Br(H \to \tau^+ \tau^-)$ can only reach up to 0.2.
Furthermore, the branching ratio $Br(H \to \tau^+ \tau^-)$ is generally lower than $Br(H \to b\bar{b})$ due to the lower mass of $\tau$ compared to $b$, as the coupling strength of Higgs with fermions is proportional to their mass.

\item In the lower left panel, the branching ratios of $H$ to light Higgs bosons are relatively small, reaching a maximum of approximately 0.6. 
Additionally, for samples where $\tan\beta$ exceeds 40, the maximum branching ratio $Br(H \to \text{light Higgs})$ is only 0.1.

\item In the lower right panel, the maximum branching ratios of $H$ to SUSY particles can approach 0.8. 
Additionally, it is observed that when $\tan\beta$ ranges from 10 to 30, the branching ratios $Br(H \to \text{SUSY})$ can approach this maximum value of 0.8, remaining above 0.5. 
When $\tan\beta$ is less than 10, the heavy Higgs $H$ predominantly decays into $t\bar{t}$; conversely, when $\tan\beta$ exceeds 30, it primarily decays into $b\bar{b}$.

\end{itemize}

\begin{figure*}[!htbp]
\centering
\includegraphics[width=1.0\textwidth]{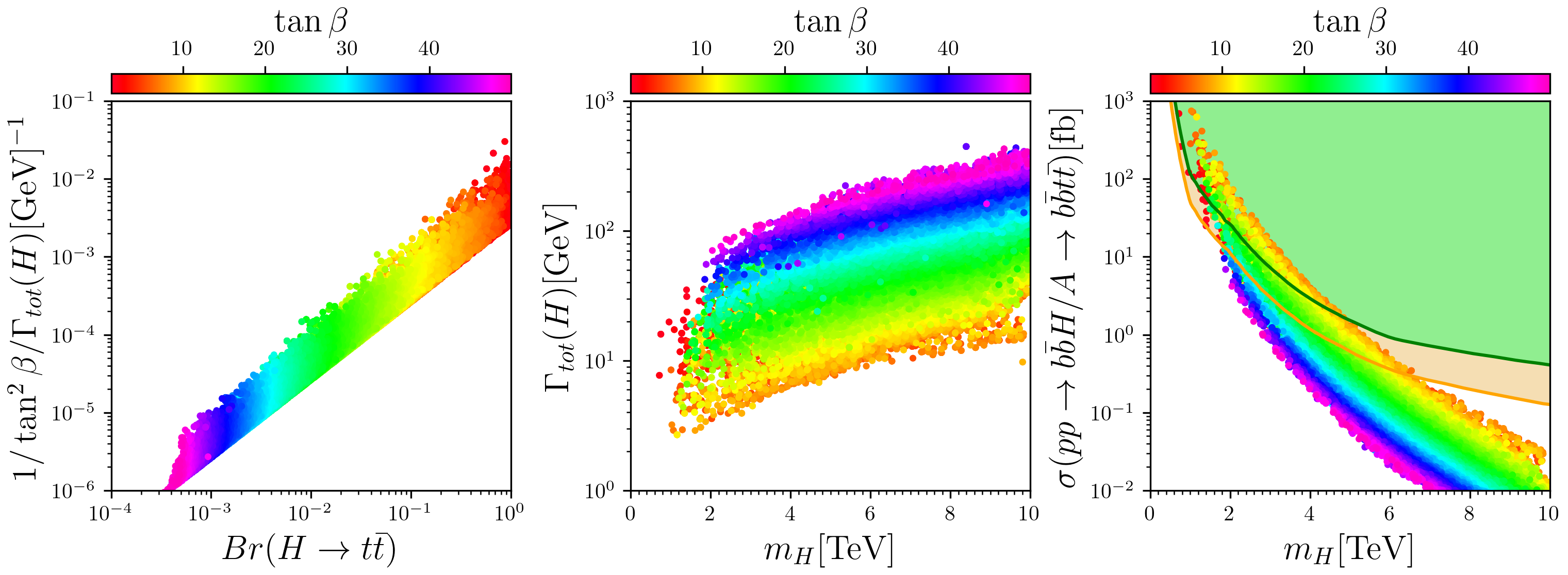}
% \vspace{-0.6cm}
\caption{
Surviving samples are shown in the planes of $1/\tan^2 \beta / \Gamma_{tot}(H)$ versus $Br(H \to t\bar{t})$ (left), heavy Higgs total decay width $\Gamma_{tot}(H)$ versus heavy Higgs mass $m_H$ (middle), and cross section of $pp \to b\bar{b}H \to b\bar{b}t\bar{t}$ versus heavy Higgs mass $m_H$ (right), where the colors of the samples indicate $\tan\beta$. 
The red and green curves represent the model-independent exclusion and discovery ranges, respectively, for the $pp \to b\bar{b}H/A \to b\bar{b}t\bar{t}$ channel, with an integrated luminosity of $3\abm$ at $100\TeV$, as taken from Fig.9 in Ref. \cite{Hajer:2015gka}.
Samples with larger values of $\tan \beta$ are plotted on top of those with smaller values.
}
\label{fig:3}
\end{figure*}

We calculate the cross sections for the process $pp \to b\bar{b} H$ in the SM with $m_H$ ranging from $0.5$ to $10\TeV$ at $\sqrt{s} = 100\TeV$ using \textsf{MG5\_aMC\_v2.6.7} \cite{Alwall:2014hca, Hirschi:2015iia}. 
The calculated cross section for $m_H$ or $m_A$ in our samples is multiplied by the square of the reduced coupling $C_{Hbb}$ and the branching ratio $Br(H/A \to t\bar{t})$. 
Since the masses and various couplings of the heavy Higgs $H$ and $A$ are very similar, along with nearly identical branching ratios $Br(H \to t\bar{t})$ and $Br(A \to t\bar{t})$, and similar reduced couplings $C_{Hbb}$ and $C_{Abb}$, the heavy Higgs bosons $H$ and $A$ are considered degenerate in the detection channel $pp \to b\bar{b}H/A \to b\bar{b}t\bar{t}$.
Therefore, the cross section for the $pp \to b\bar{b}H/A \to b\bar{b}t\bar{t}$ channel is twice that of the individual $H$ or $A$ channels.
Production rates for our samples in the $pp \to b\bar{b}H/A \to b\bar{b}t\bar{t}$ channel are presented in the left panel of Fig. \ref{fig:3}, where colors indicate $\tan\beta$. 
The red and green curves represent the model-independent exclusion and discovery reaches, respectively, with an integrated luminosity of $3\abm$ at $100\TeV$, as depicted in Fig.9 of Ref. \cite{Hajer:2015gka}. 
In calculating the SM cross section, we employed both the four-flavor scheme (4FS) and five-flavor scheme (5FS) cross sections \cite{Wiesemann:2014ioa}, and combined them using the formula from Ref. \cite{Harlander:2011aa}:
\begin{equation}
	\sigma = \frac{\sigma^{\rm 4FS} + \omega \sigma^{\rm 5FS}}{1+\omega},
\end{equation}
where $\omega = \ln(m_H/m_b) - 2$.

In Fig. \ref{fig:3} we show the cross section for the $pp \to b\bar{b}H/A \to b\bar{b}t\bar{t}$ channel of the CP-even doublet-dominated heavy Higgs $H/A$ in the scNMSSM, with colors representing $\tan \beta$. 
Since the heavy Higgs $A$ and $H$ are considered degenerate, the cross section $\sigma(pp \to b\bar{b}H/A \to b\bar{b}t\bar{t}) = 2\sigma(pp \to b\bar{b}H \to b\bar{b}t\bar{t}) $. 
The following observations can be drawn from these plots:
\begin{itemize}
\item In the left panel, $1/\tan^2 \beta / \Gamma_{tot}(H)$ appears to be directly proportional to $Br(H \to t\bar{t})$. 
This is because the decay diagram for $Br(H \to t\bar{t})$ includes a coupling vertex $C_{Huu}$, and when calculating the decay cross-section, a $C_{Huu}^2$ term is introduced. 
Thus, $Br(H \to t\bar{t})$ is proportional to $1/\tan^2 \beta$. Additionally, the branching ratio $Br(H \to t\bar{t})$ is inversely proportional to the total decay width $\Gamma_{tot}(H)$, which is represented as:
\begin{align}
    Br(H \to t\bar{t}) = \frac{\sigma(H \to t\bar{t})}{\Gamma_{tot}(H)} \sim  \frac{1/\tan^2\beta}{\Gamma_{tot}(H)} \, .
\end{align}

\item In the middle panel, the total decay width $\Gamma_{tot}(H)$ of the heavy Higgs increases exponentially with $\tan\beta$. 
Additionally, when $\tan\beta$ remains constant, $\Gamma_{tot}(H)$ increases with the mass of the heavy Higgs $m_H$.

\item In the right panel, the cross section $\sigma(pp \to b\bar{b}H/A \to b\bar{b}t\bar{t})$ decreases rapidly as the mass of the heavy Higgs $m_H$ increases. The cross section for $pp \to b\bar{b}H \to b\bar{b}t\bar{t}$ can be approximated as follows:
\begin{align}
& \sigma(pp \to b\bar{b}H \to b\bar{b}t\bar{t}) \nonumber \\
    \approx & \sigma(pp \to b\bar{b}H^{SM}) \cdot C_{Hdd}^2 \cdot Br(H \to t\bar{t}) \nonumber \\
    % \approx & \sigma(pp \to b\bar{b}H^{SM}) \cdot \tan^2\beta \cdot \frac{1/\tan^2\beta}{\Gamma_{tot}(H)} \nonumber \\
    \approx &  \frac{\sigma(pp \to b\bar{b}H^{SM})}{\Gamma_{tot}(H)}
\end{align}
This decline is because $\sigma(pp \to b\bar{b}H \to b\bar{b}t\bar{t})$ is proportional to $\sigma(pp \to b\bar{b}H^{SM})$, and the production cross section $\sigma(pp \to b\bar{b}H^{SM})$ diminishes with an increase in mass.
It can also be observed that samples with smaller $\tan\beta$ values have larger cross sections $\sigma(pp \to b\bar{b}H \to b\bar{b}t\bar{t})$. 
This is because $\sigma(pp \to b\bar{b}H \to b\bar{b}t\bar{t})$ is inversely proportional to the total decay width $\Gamma_{tot}(H)$, and $\Gamma_{tot}(H)$ exponentially increases as $\tan\beta$ increases.

\item In the right panel, the regions above the green and red curves indicate where the samples can be covered by 2 $\sigma$ and 5 $\sigma$, respectively, with an integrated luminosity of $3 \abm$ at $100 \TeV$. 
This implies that through the $pp \to b\bar{b}H/A \to b\bar{b}t\bar{t}$ channel in the scNMSSM, samples with a heavy Higgs mass $m_H < 2 \TeV$ can be tested at at the $100\TeV$ collider with $3 \abm$ of integrated luminosity. 
For samples with the heavy Higgs mass $m_H > 7 \TeV$, they fall below the exclusion and discovery curves, thus they cannot be discovered or excluded. 
Samples with the heavy Higgs mass in the range of $2-7 \TeV$ and $\tan\beta < 20$ can be tested at the $100\TeV$ collider with $3 \abm$ of integrated luminosity.

\end{itemize}

% \usepackage{booktabs}
\begin{table}[]
\centering
\caption{\label{tab:1}Four Benchmark Points for Surviving Samples, where $\sigma(X)$ denotes the cross section $\sigma(pp \to b\bar{b}X \to b\bar{b}t\bar{t})$. 
Here, $H$ and $A$ represent the doublet-dominated heavy Higgs bosons, while $S_{33}^2$ and $P_{22}^2$ indicate the singlet components in $H$ and $A$ respectively.}
\begin{tabular}{@{}ccccc@{}}
\toprule
                                                    & P1      & P2      & P3      & P4      \\ \midrule
$\lambda$                                           & 0.61    & 0.21    & 0.10    & 0.10    \\
$\kappa$                                            & 0.36    & -0.21   & -0.42   & 0.67    \\
$\tan\beta$                                         & 2.07    & 4.98    & 20.02   & 47.91   \\
$\mu[\rm GeV]$                                               & 361     & 345     & 295     & 498     \\
$M_0[\rm GeV]$                                               & 8072    & 1506    & 5811    & 9596    \\
$M_{12}[\rm GeV]$                                            & 3402    & 5569    & 3017    & 9289    \\
$A_0[\rm GeV]$                                               & 7306    & -6275   & 539     & 9862    \\
$A_\lambda[\rm GeV]$                                         & 4961    & 1230    & 4983    & 2357    \\
$A_\kappa[\rm GeV]$                                          & 2720    & 252     & 3769    & -3588   \\
$m_{h_1}[\rm GeV]$                                           & 124     & 124     & 124     & 125     \\
$m_{h_2}[\rm GeV]$                                           & 341     & 648     & 1942    & 6796    \\
$m_{H}[\rm GeV]$                                           & 720     & 2025    & 4030    & 7950    \\
$m_{a_1}[\rm GeV]$                                           & 512     & 277     & 2920    & 1860    \\
$m_{A}[\rm GeV]$                                           & 716     & 2026    & 4031    & 7950    \\
% $m_{H^\pm}$                                         & 713     & 2026    & 4031    & 7950    \\
$S_{31}^2$                                            & 0.99    & 1.00    & 1.00    & 1.00    \\
$S_{32}^2$                                            & 0.00    & 0.00    & 0.00    & 0.00    \\
$S_{33}^2$                                            & 0.01    & 0.00    & 0.00    & 0.00    \\
$P_{21}^2$                                            & 1.00    & 1.00    & 1.00    & 1.00    \\
$P_{22}^2$                                            & 0.00    & 0.00    & 0.00    & 0.00    \\
% $C_{h_1uu}$                                         & 1.0     & 1.0     & 1.0     & 1.0     \\
$C_{h_2uu}$                                         & 0.1     & 0.0     & 0.0     & 0.0     \\
$C_{Huu}$                                         & -0.5    & -0.2    & -0.1    & 0.0     \\
$C_{a_1uu}$                                         & 0.0     & 0.0     & 0.0     & 0.0     \\
$C_{Auu}$                                         & 0.5     & 0.2     & 0.0     & 0.0     \\
% $C_{h_1dd}$                                         & 1.0     & 1.0     & 1.0     & 1.0     \\
$C_{h_2dd}$                                         & 0.4     & 0.1     & 0.0     & 0.3     \\
$C_{Hdd}$                                         & 2.0     & 5.0     & 20.0    & 47.9    \\
$C_{a_1dd}$                                         & 0.0     & -0.2    & -0.3    & 0.1     \\
$C_{Add}$                                         & 2.1     & 5.0     & 20.0    & 47.9    \\
% $Br(h_1\to t\bar{t})$                               & 0       & 0       & 0       & 0       \\
$Br(h_2\to t\bar{t})$                               & 0       & 0.01    & $4.4 \times 10^{-4}$ & $7.9 \times 10^{-8}$ \\
$Br(H\to t\bar{t})$                               & 0.87    & 0.41    & 0.01    & $5.7 \times 10^{-4}$ \\
$Br(a_1\to t\bar{t})$                               & 0.95    & 0       & $6.0 \times 10^{-5}$ & $2.6 \times 10^{-7}$ \\
$Br(A\to t\bar{t})$                               & 0.91    & 0.41    & 0.01    & $5.7 \times 10^{-4}$ \\
% $\Gamma_{tot}(h_1)[\rm GeV]$                                 & 3.8 \times 10^{-3} & 4.0 \times 10^{-3} & 0.00    & 0.00    \\
$\Gamma_{tot}(h_2)[\rm GeV]$                                 & 0.20    & 0.39    & 0.65    & 59.04   \\
$\Gamma_{tot}(H)[\rm GeV]$                                 & 7.71    & 8.99    & 45.27   & 236.25  \\
$\Gamma_{tot}(a_1)[\rm GeV]$                                 & $4.8 \times 10^{-4}$ & $1.2 \times 10^{-4}$ & 1.08    & 0.59    \\
$\Gamma_{tot}(A)[\rm GeV]$                                 & 8.93    & 9.12    & 45.31   & 236.33  \\
$\sigma(h_2)[\rm fb]$ & 0       & 0.01    & $1.6 \times 10^{-6}$ & $4.3 \times 10^{-11}$ \\
$\sigma(H)[\rm fb]$ & 331.80  & 16.80   & 0.29    & $2.9 \times 10^{-3}$ \\
$\sigma(a_1)[\rm fb]$ & 0.10    & 0.00    & $1.7 \times 10^{-6}$ & $5.8 \times 10^{-9}$ \\
$\sigma(A)[\rm fb]$ & 360.03  & 16.73   & 0.29    & $2.9 \times 10^{-3}$ \\ \bottomrule
\end{tabular}
\end{table}

In Table~\ref{tab:1}, we present four benchmark samples detailing the Higgs sector, where $\sigma(X)$ represents the cross section $\sigma(pp \to b\bar{b}X \to b\bar{b}t\bar{t})$.
The heavy Higgs bosons $H$ and $A$, corresponding to $h_3$ and $a_2$ respectively, are doublet-dominated, while $S_{33}^2$ and $P_{22}^2$ indicate the singlet components in $H$ and $A$. 
We find that $H$ and $A$ have minimal singlet components, suggesting that $h_2$ and $a_1$ are primarily singlet-dominated. 
Due to their weak coupling to fermions, these singlet-dominated bosons, $h_2$ and $a_1$, are difficult to detect at colliders.

\section{Conclusion}
\label{sec:con}

In this study, we explore the potential for detecting heavy Higgs bosons in the $pp \to b\bar{b}H/A \to b\bar{b}t\bar{t}$ channel at a 100 TeV hadron collider within the semi-constrained NMSSM. 
First, we scan the relevant parameter space with the \textsf{NMSSMTools} package, which includes theoretical constraints such as vacuum stability and Landau poles, as well as experimental constraints like Higgs data, B physics, sparticle searches, dark matter relic density, and direct detection experiments. 
We observe that singlet-dominated Higgs bosons $S$ are difficult to detect due to their limited interactions outside the Higgs sector. Therefore, our analysis primarily focuses on the more detectable heavy, doublet-dominated CP-even Higgs $H$ and CP-odd Higgs $A$, limiting their masses to below $10\TeV$ to remain detectable.
The presence of a CP-even Higgs ($h_1$ or $h_2$) resembling the 125 GeV SM-like Higgs does not affect these findings. 
Since that the heavy Higgs $H$ and $A$ are nearly identical in mass and couplings, the cross section for the combined channel $pp \to b\bar{b}H/A \to b\bar{b}t\bar{t}$ is effectively double that of the single $H$ channel.

We calculated their decay branching ratios and production rates, and compared these with the simulation results cited in Ref. \cite{Hajer:2015gka}. 
Finally, we draw the following conclusions about the heavy Higgs bosons $A$ and $H$, with masses ranging from 0.6 to 10 TeV, in the semi-constrained NMSSM:
\begin{itemize}
\item When the heavy Higgs bosons are doublet-dominated, their reduced couplings with up-type fermions, $C_{Huu}$ and $C_{Auu}$, are approximately equal to $1/\tan \beta$. 
This relationship causes the branching ratio of $H/A$ to $t\bar{t}$ to be inversely proportional to $\tan\beta$; specifically, a smaller $\tan\beta$ results in a larger branching ratio $Br(H \to t\bar{t})$. 
Conversely, the reduced couplings with down-type fermions, $C_{Hdd}$ and $C_{Add}$, approximate to $\tan \beta$, leading to branching ratios of $H$ to $b\bar{b}$ and $\tau^+ \tau^-$ that are directly proportional to $\tan\beta$.

\item When $\tan\beta$ is less than 10, the heavy Higgs $H$ predominantly decays into $t\bar{t}$, with the branching ratio $Br(H \to t\bar{t})$ reaching up to 1. 
When $\tan\beta$ exceeds 30, it primarily decays into $b\bar{b}$, with the branching ratio $Br(H \to b\bar{b})$ up to 0.8 and $Br(H \to \tau^+ \tau^-)$ also reaching up to 0.2. 
For $\tan\beta$ values between 10 and 30, the branching ratio $Br(H \to SUSY)$ is dominant, reaching up to 0.8.

\item The branching ratio $Br(H \to t\bar{t})$ is proportional to $1/\tan^2 \beta$ and inversely proportional to the total decay width $\Gamma_{\text{tot}}(H)$. 
Furthermore, the total decay width of the heavy Higgs, $\Gamma_{\text{tot}}(H)$, increases exponentially with $\tan\beta$.

\item The cross section $\sigma(pp \to b\bar{b}H/A \to b\bar{b}t\bar{t})$ decreases rapidly as the mass of the heavy Higgs ($m_H$) increases, and it is inversely proportional to the total decay width $\Gamma_{\text{tot}}(H)$.
Consequently, this cross section also decreases exponentially with increasing $\tan\beta$.

\item For the $pp \to b\bar{b}H/A \to b\bar{b}t\bar{t}$ channel at a 100 TeV collider with 3 ab$^{-1}$ of integrated luminosity in the semi-constrained NMSSM: 
\begin{itemize}
    \item Heavy Higgs bosons with a mass $m_H < 2$ TeV can be tested.
    \item Heavy Higgs bosons with a mass $m_H > 7$ TeV fall below the exclusion and discovery thresholds, and therefore cannot be discovered or excluded.
    \item For heavy Higgs masses in the range of 2-7 TeV, those with $\tan\beta \lesssim 20$ can be tested, while those with $\tan\beta \gtrsim 20$ cannot be discovered or excluded.
\end{itemize}

% \item For the $pp \to b\bar{b}H/A \to b\bar{b}t\bar{t}$ channel at a 100 TeV collider with 3 ab$^{-1}$ of integrated luminosity in the semi-constrained NMSSM, the detectability of heavy Higgs bosons varies by mass and $\tan\beta$. 
% Heavy Higgs bosons with a mass less than 2 TeV can be tested. 
% In contrast, those with a mass greater than 7 TeV fall below the exclusion and discovery thresholds, making them undetectable. 
% For masses between 2 and 7 TeV, Heavy Higgs bosons with $\tan\beta$ less than 20 can be tested, while those with $\tan\beta$ greater than 20 are beyond the current capabilities for discovery or exclusion.

\end{itemize}

\begin{acknowledgments}
This work was supported by the National Natural Science Foundation of China (NNSFC) under
grant Nos. 12275066 and 11605123.
\end{acknowledgments}

\appendix

% \section{Appendixes}

% To start the appendixes, use the \verb+\appendix+ command.
% This signals that all following section commands refer to appendixes
% instead of regular sections. Therefore, the \verb+\appendix+ command
% should be used only once---to setup the section commands to act as
% appendixes. Thereafter normal section commands are used. The heading
% for a section can be left empty. 

% Note the equation numbers in this appendix, produced with the
% subequations environment:
% \begin{subequations}
% \begin{eqnarray}
% E&=&mc, \label{appa}
% \\
% E&=&mc^2, \label{appb}
% \\
% E&\agt& mc^3. \label{appc}
% \end{eqnarray}
% \end{subequations}
% They turn out to be Eqs.~(\ref{appa}), (\ref{appb}), and (\ref{appc}).

% \nocite{*}

% \bibliographystyle{apsrev4-2}
\bibliographystyle{apsrev4-1}
\bibliography{apssamp}% Produces the bibliography via BibTeX.

\end{document}